\documentclass[a4paper]{article}

\usepackage{icrc2013}

\def\simleq{\; \raise0.3ex\hbox{$<$\kern-0.75em \raise-1.1ex\hbox{$\sim$}}\; }
\def\simgeq{\; \raise0.3ex\hbox{$>$\kern-0.75em \raise-1.1ex\hbox{$\sim$}}\; }

\newcommand{\GeV}{{\rm GeV}}
\newcommand{\GV}{{\rm GV}}
\newcommand{\TeV}{{\rm TeV}}

\newcommand{\erg}{{\rm erg}}

\newcommand{\kpc}{{\rm kpc}}

\newcommand{\cm}{{\rm cm}}

\newcommand{\s}{{\rm s}}

\title{Three dimensional modeling of CR propagation}

\shorttitle{3D modeling of CR propagation}

\authors{D. Gaggero$^{1,2}$, L. Maccione$^{3,4}$, G. Di Bernardo$^{5,6}$, C. Evoli$^{7}$, and D. Grasso$^{8}$ 
}

\afiliations{
$^1$SISSA, via Bonomea 265, I-34136, Trieste, Italy\\ 
$^2$INFN, sezione di Trieste, via Valerio 2, I-34127, Trieste, Italy\\
$^3$Ludwig-Maximilians-Universit\"{a}t, Theresienstra{\ss}e 37, D-80333 M\"{u}nchen, Germany\\
$^4$Max-Planck-Institut f\"{u}r Physik (Werner Heisenberg Institut), F\"{o}hringer Ring 6, D-80805 M\"{u}nchen, Germany\\	
$^5$Department of Physics, University of Gothenburg, SE 412 96 Gothenburg, Sweden\\
$^6$Department of Astronomy and Theoretical Astrophysics Center, University of California Berkeley, Berkeley, CA 94720\\
$^7$Institut f\"ur Theoretische Physik, Universit\"{a}t Hamburg, Luruper Chaussee 149, D-22761 Hamburg, Germany\\
$^8$INFN, sezione di Pisa, Largo B. Pontecorvo 3, I-56127 Pisa, Italy
}

\email{daniele.gaggero@sissa.it}

\abstract{We present here a major upgrade of {\tt DRAGON}, a numerical package that computes the propagation of a wide set of CR species from both astrophysical and exotic origin in the Galaxy in a wide energy range from tens of MeV to tens TeV. {\tt DRAGON} takes into account all relevant processes in particular diffusion, convection, reacceleration, fragmentation and energy losses. 
For the first time, we present a full 3D version of {\tt DRAGON} with anisotropic position-dependent diffusion. In this version, the propagation is calculated within a 3D cartesian grid and the user is able to implement realistic and structured three dimensional source, gas and regular magnetic field distributions. Moreover, it is possible to specify an arbitrary function of position and rigidity for the diffusion coefficients in the parallel and perpendicular direction to the regular magnetic field of the Galaxy. The code opens many new possibilities in the study of CR physics. In particular, we can study for the first time the impact of the spiral arm structure on the leptonic spectra: taking into account the fact that we live in an interarm region, far from most sources, we obtain -- due to increased energy losses -- a steeper electron spectrum compared to the assumption of a smooth source term. We discuss the implications of these results on our understanding on leptonic spectra and we briefly mention future studies that can be performed with our new  3D code.}

\keywords{high-energy cosmic rays; 3D propagation; numerical codes.}

\begin{document}
\maketitle


\section{Introduction}

In the latest years the measurements of the CR leptonic fluxes have reached an unprecedented accuracy.
Thanks to experiments such as Fermi-LAT, PAMELA, AMS-02 we have now high-statistic data on the electron+positron (CRE) flux up to $\simeq 1$ TeV and on the positron fraction (PF) up to $\simeq 350$ GeV.
With increasing energy, the leptonic momentum losses (due to the well-known phenomena of Inverse Compton scattering off diffuse light and synchrotron emission) become more and more effective, so in the TeV region the horizon is limited to less than 1 kpc. 
For this reason, as the energy increases, the accurate datasets recently collected probe smaller and smaller regions, and the details of large, medium and small scale structures in the Galactic environment may play a major role in their interpretation.
Therefore, it is important to implement models of CR production and propagation where inhomogeneities in the source and gas distributions and a more detailed description of diffusion are present.
On the modeling side, most simulations performed so far include a very simple description of the Galaxy, in which azimuthal symmetry is assumed, CR sources follow a very smooth spatial distribution $Q(R,z)$ and the diffusion coefficient is a scalar. In this way, no difference between transport along the parallel and perpendicular directions with respect to the regular magnetic field is taken into account. 

This description is not satisfactory because it is well known that $D_\parallel$ and $D_\bot$ have opposite behaviors with the turbulence level \cite{DeMarco:2007eh}. Moreover, the Galaxy has a spiral arm structure with the Solar System lying in a local overdensity within a wide interarm region. On even more local scale, the interstellar medium shows local overdensities and large voids, and structures as the Local Bubble may play a role in the transport of high energy leptons.

Here we present a new version of our CR propagation code {\tt DRAGON} \cite{dragon1,dragon2,dragon3} in which full 3D CR propagation is implemented in cartesian coordinates. In this new version of the code the user is allowed to use an arbitrary 3D distribution for the gas and source distributions, a realistic model for the regular magnetic field and arbitrary functions of position and rigidity for the parallel and perpendicular diffusion coefficients. 
These features allow one to explore a wide class of phenomena largely neglected so far. After a brief technical description of the code and some details about its testing phase, we describe a relevant result that we have obtained with a 3D propagation model and point out possible future developments.

\section{The 3D anisotropic version of {\tt DRAGON}}

\subsection{The equation}

The most general three dimensional equation for CR diffusion in cartesian coordinates may be written in this form:

\begin{eqnarray}
\frac{\partial f}{\partial t}\,=\,Q &+& \alpha_{xx}\partial_x^2 f + \alpha_{yy}\partial_y^2 f + \alpha_{zz}\partial_z^2 f\nonumber\\
&+& 2 \delta_{xy}\partial_x \partial_y f + 2 \delta_{xz}\partial_x \partial_z f + 2 \delta_{yz}\partial_y \partial_z f \nonumber\\
&+& u_x \partial_x f + u_y \partial_y f + u_z \partial_z f 
\end{eqnarray}

In this equation the following position-dependent coefficients appear next to the spatial derivatives:

\begin{eqnarray*}
\alpha_{xx}(x,y,z) \,&=&\,\, (D_{\parallel} - D_{\bot}) b_{x}^2 \,+\, D_{\bot}\\
\alpha_{yy}(x,y,z) \,&=&\,\, (D_{\parallel} - D_{\bot}) b_{y}^2 \,+\, D_{\bot}\\
\alpha_{zz}(x,y,z) \,&=&\,\, (D_{\parallel} - D_{\bot}) b_{z}^2 \,+\, D_{\bot}\\
\delta_{xy}(x,y,z) \,&=&\,\, (D_{\parallel} - D_{\bot}) b_{x}b_{y} \,+\, D_{\bot}\\
\delta_{xz}(x,y,z) \,&=&\,\, (D_{\parallel} - D_{\bot}) b_{x}b_{z} \,+\, D_{\bot}\\
\delta_{yz}(x,y,z) \,&=&\,\, (D_{\parallel} - D_{\bot}) b_{y}b_{z} \,+\, D_{\bot}\\
u_x(x,y,z) \,&=&\,\, \partial_x \alpha_{xx} \,+\, \partial_y \delta_{xy} \,+\, \partial_z \delta_{xz}\\
u_y(x,y,z) \,&=&\,\, \partial_x \delta_{xy} \,+\, \partial_y \alpha_{yy} \,+\, \partial_z \delta_{yz}\\
u_z(x,y,z) \,&=&\,\, \partial_x \delta_{xz} \,+\, \partial_y \delta_{yz} \,+\, \partial_z \alpha_{zz}
\end{eqnarray*}

where: 
\begin {itemize}
\item $b_x(x,y,z)$, $b_y(x,y,z)$ and $b_z(x,y,z)$ are the versors of the regular magnetic field
\item $D_{\parallel}(x,y,z,E)$ and $D_{\bot}(x,y,z,E)$ are the diffusion coefficients in the parallel and perpendicular directions with respect to the regular magnetic field
\end{itemize}

The user is allowed to choose his own preferred expression for $D_{\parallel}(\vec{x},E)$ and $D_{\bot}(\vec{x},E)$; the only limitation is separability in space and energy dependence.

\subsection{Testing the code: Green's functions}

In this Section we briefly describe, for illustrative purposes, some relevant Green's function tests.

%
\begin{figure}[!h]
  \centering
  \includegraphics[width=3.in]{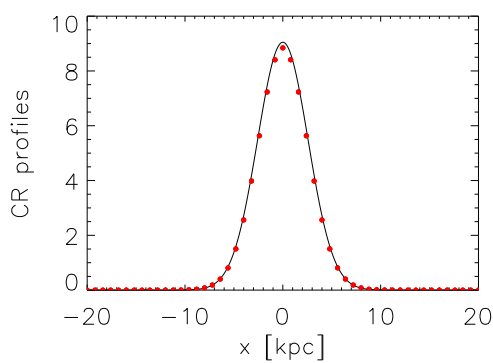}
  \caption{Comparison with analytical solution after 10 timesteps; profile along $x$ axis, $y = z = 0$; parallel diffusion is dominant.}
  \label{fig:test_dpar_2}
\end{figure}

%
\begin{figure}[!h]
  \centering
  \includegraphics[width=3.in]{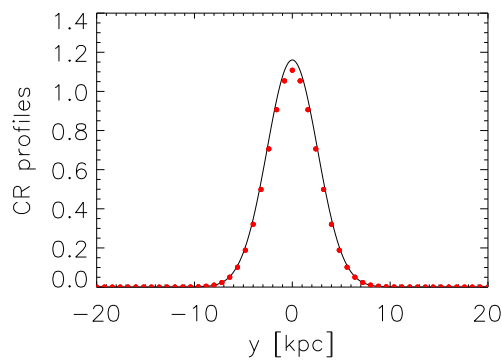}
  \caption{Comparison with analytical solution  after 10 timesteps; profile along $y$ axis, $x = z = 0$. Perpendicular diffusion is dominant. 
  }
  \label{fig:test_dperp_2}
\end{figure}

  We considered a toy model for the regular magnetic field ($\vec{B}\propto\hat{x}$ ante constant) and the following extreme cases:

  \begin{itemize}
   \item $D_{\parallel} \,>>\, D_{\bot}$, with both coefficients constant in space
   \item $D_{\parallel} \,<<\, D_{\bot}$, with both coefficients constant in space  
  \end{itemize}

  We initialized the CR distribution to a Dirac delta centered in the origin of the coordinate system and we compared the numerical solution calculated with {\tt DRAGON 3D} to the analytical solution.

  \begin{itemize}
   \item 
	In the first case the analytical solution if the 1D Green's function for pure diffusion along $x$ reads:
	\begin{equation}
	G(x,t) = \frac{1}{\sqrt{4\pi Dt}}\, \exp{\left[\frac{-x^2}{4\, D\, t}\right]}
	\end{equation}
	We set the following parameters: $\Delta t = 0.1 \,{\rm Myr}$,  $D_{\parallel} = 10^{30} \,{\rm cm^2 s^{-1}}$ ($\sqrt{D\Delta t} \simeq 500\, {\rm pc}$); $\Delta x = 800 \,{\rm pc}$; 
	diffusion box in $x$ direction: $-40 < x < 40 \,{\rm kpc}$.
	The reader can see in Fig.~\ref{fig:test_dpar_2} the comparison between numerical and analytical solution after 10 time-steps.
   \item 
	In the second case the analytical solution if the 2D Green function for pure diffusion along $y,\,z$ is:
	\begin{equation}
	G(y,z,t) = \frac{1}{\sqrt{(4\pi Dt)^2}}\, \exp{\left[\frac{-(y^2 + z^2)}{4\, D\, t}\right]}
	\end{equation}
	Notice the different scaling with time of the normalization of the Gaussian.
	We set the following parameters: $\Delta t = 0.1 \,{\rm Myr}$,  $D_{\bot} = 10^{30} \,{\rm cm^2 s^{-1}}$ ($\sqrt{D\Delta t} \simeq 500\, {\rm pc}$); $\Delta x = 800 \,{\rm pc}$; 
	diffusion box in $y,\, z$ direction: $-40 < x < 40 \,{\rm kpc}$.
	We plotted in Fig.~\ref{fig:test_dperp_2} the comparison between numerical and analytical solution. We have a good match in this case too. 
  \end{itemize}

\section{Impact of a realistic 3D source function on the CR leptonic spectra}

It is well known that, in order to reproduce the CR $e^-$ and $e^+$ spectra, it is necessary to consider -- beyond a conventional component of primary electrons and secondary electrons and positrons -- some extra contribution of unclear origin. 

In this class of models a very steep injection for the primary component is required to match the data: the injection slope is $-2.65 	\div -2.70$ depending on the diffusion setup and appears to be in strong tension with:
\begin{itemize}
\item that inferred from radio observations of SNRs, $\langle\gamma\rangle=2.0\pm0.3$ \cite{Delahaye:2010ji}. 
\item the values $2.2 \div2.4$ required to reproduce the CR nuclei spectra. 
\item the shock acceleration theory which generally predicts the same spectral index, close to $2\div2.3$, for electrons and nuclei \cite{Caprioli:2011ze}
\end{itemize}

\begin{figure}[!h]
  \centering
  \includegraphics[width=3.in]{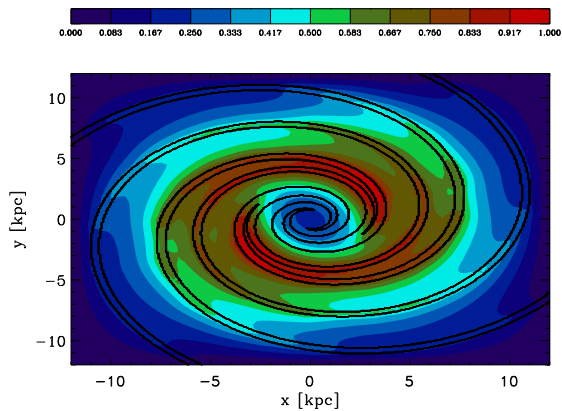}
  \caption{CR electron distribution on the Galactic plane at 1 GeV.}
  \label{fig:electrons_faceon_1}
\end{figure}

\begin{figure}[!h]
  \centering
  \includegraphics[width=3.in]{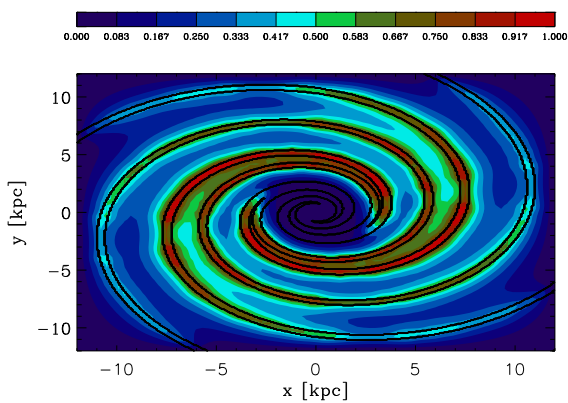}
  \caption{CR electron distribution on the Galactic plane at 100 GeV.}
  \label{fig:electrons_faceon_100}
\end{figure}

Here instead we present, as an important application of our 3D code, a realistic model in which this serious problem is naturally solved.
The idea is to consider the fact that we live in an interarm region and the bulk of the CR sources are expected to lay in the arms: the energy losses suffered by the $e^{\pm}$ are highly enhanced with respect to the simple model in which CR sources are smoothly distributed in the Galaxy, because of the greater average distance that $e^{\pm}$ have to cross to arrive at Earth. This effect allows us to fit the observed CRE spectra and PF, including the new data recently provided by AMS-02, by adopting a CRE injection index very close to the one needed for CR protons and other nuclear species. 

\begin{figure}[!h]
  \centering
  \includegraphics[width=3.in]{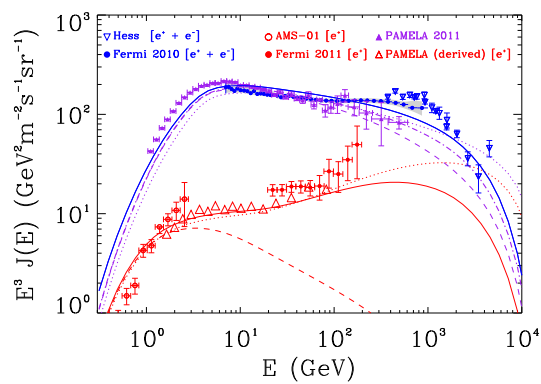}
  \caption{The $e^- + e^+$ (blue solid), $e^-$ (purple dot-dashed) and $e^+$ (red solid) propagated spectra computed under the hypothesis that the background and extra-component sources are distributed in the spiral arm as shown Fig.~\ref{fig:electrons_faceon_100}.  
  Dashed lines represent the background contributions to the $e^-$ (purple) and $e^+$ red spectra. The dotted lines represent the spectra obtained for a continuos (no spiral arms) source distribution.  PAMELA $e^+$ data have been derived (without errors) starting from the PF and $e^-$ spectrum released by the same collaboration.}
  \label{fig2}
\end{figure}

\begin{figure}[!h]
  \centering
  \includegraphics[width=3.in]{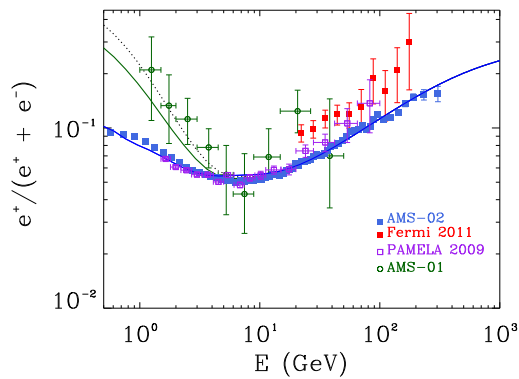}
  \caption{The PF computed assuming the extra-component sources are located only in spiral arms. The solid and dashed lines correspond to the AMS-02 and AMS-01 data taking periods respectively. The dotted line is the interstellar PF.}
  \label{fig3}
\end{figure}

\begin{figure}[!h]
  \centering
  \includegraphics[width=3.in]{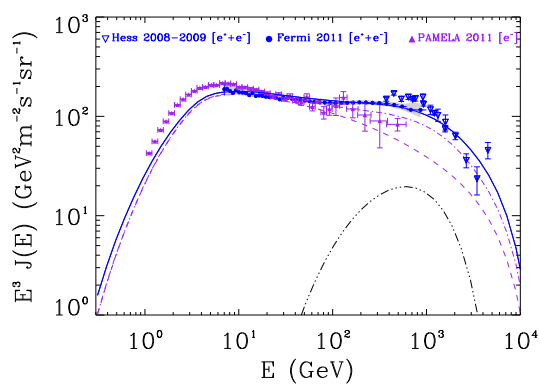}
  \caption{The $e^- + e^+$, and $e^-$ propagated spectra computed assuming the extra-component sources are located only in spiral arms and that one nearby electron accelerator is present.  The triple-dot-dashed curve show the contribution of a local source at the position of the Vela SNR.}
  \label{fig4}
\end{figure}

In our model both the conventional CR sources and the extra electron+positron source term are located in the arms. 
For all nuclear species we use an unbroken power-law source spectrum with the same spectral index $\gamma_{0, p} = 2.28$.   
For the $e^-$ conventional source spectrum we assume a broken power-law as required to consistently reproduce the spectrum of the diffuse radio emission of the Galaxy and high energy CR data \cite{DiBernardo:2012zu}: below $4~\GeV$ we adopt a spectral index $1.2$ while above that energy we tune it against PAMELA data (see below). 
For the extra-component source spectrum we assume a simple power-law with exponential cutoff: $J_{EC}(e^\pm) \propto E^{-\gamma_{0, {\rm EC}}}  \exp({-E/E_{\rm cut}})$.
The CR propagation is treated in 3D isotropic mode, without reacceleration and convection. The diffusion coefficient has the following dependence on the rigidity
 $\rho$: $D(\rho) = \beta^{-0.4}~ D_0 \left( \rho/\rho_0 \right)^\delta$, with $\rho_{0}=3~\GV$ 
The propagation parameters are chosen as follows: $D_{0}=3\times10^{28}~\cm^{2}/\s$, $\delta=0.6$, half-halo height of $4~\kpc$ and no reacceleration. These parameters are tuned to fit the B/C ratio as well as other light nuclei ratios. The proton source term and injection is tuned to reproduce the PAMELA dataset.
Concerning low energies, we consider the effect of solar modulation using the recently developed {\tt HelioProp} numerical code: this package \cite{Maccione:2012cu} solves the CR transport equation in the heliosphere accounting for charge-dependent drifts.

The effect of energy losses in such a scenario is evident in Fig.~\ref{fig:electrons_faceon_1} and Fig.~\ref{fig:electrons_faceon_100} where it is clear that high energy electrons (100~GeV in our examples) can be found in the vicinity of the sources only, and are strongly suppressed in the interarm regions where the Solar System is located. 
The spiral arm pattern used here is taken from \cite{Blasi:2011fi}.
The main result can be seen in Fig.~\ref{fig2} and Fig.~\ref{fig3}: with an injection index $\gamma_{0,e} = 2.38$ above the break, the spiral arm pattern induces a steepening that leaves room for an extra component and permits to reproduce both PAMELA electrons and AMS-02 PF. 
This injection index is closer to that used for nuclei and compatible with shock acceleration theory. 
For the extra component we use an injection index $\gamma_{0,e (extra)} = 1.7$ 
and a 10 TeV cutoff.

Without the spiral pattern a value of $2.38$ for the injection index of the conventional component would not be compatible with the data since the propagated spectrum would be too hard and a relevant contribution from an extra source (needed to fit AMS) would overshoot the electron data: this can be clearly seen in Fig.~\ref{fig2} (dotted lines).
We remark that low-energy data regarding the PF taken at different times from different experiments are correctly reproduced (see Fig.~\ref{fig3}) adopting the correct heliospheric parameters (mainly tilt angle of the current sheet and polarity, see \cite{Maccione:2012cu}) that correspond to the data taking period and tuning the heliosperic diffusion coefficient to the data.

We also point out that, at high energy (above $\sim200$~GeV), the CRE spectrum measured by Fermi-LAT is not well reproduced (see Fig.~\ref{fig2}). The model prediction is lower than the measurement. This is not surprising since at those energies the effect of local sources may be important. Nevertheless, we do not expect many different contributions since the propagation in the local ISM (within 1~kpc) is expected to be highly anisotropic and should take place along streams \cite{Kistler:2012ag} with low probability of intersecting the Solar System. We checked, for illustrative purposes, that a single electron accelerator with energy output $\simeq 3.6 \cdot 10^{47}~\erg$ located nearby (d = $290$~pc, a position compatible with the Vela SNR) yields a good combined fit of Fermi-LAT and AMS-02 datasets in the whole energy range, adopting a spectral injection index  $\gamma=2.1$ and a cutoff $E_{cut} = 1~\TeV$ (see Fig.~\ref{fig4}).

\section{Discussion and conclusions}

We presented a major update of {\tt DRAGON}. 
This version of the code allows one to compute CR propagation adopting a realistic 3D model of the Galaxy in which the user is free to choose three-dimensional arbitrary source functions, gas distributions, magnetic field models and different position-dependent models for diffusion in the parallel and perpendicular directions with respect to the Galactic magnetic field.

We have tested the code by reproducing the Green's function for parallel and perpendicular propagation. 
Then, we have presented an important result we obtained in the isotropic diffusion mode and taking into account for the first time a realistic 3D structure of CR sources. If the sources are located mainly in the spiral arms, and given that the Sun is located in an interarm region, the enhanced energy losses permit to reproduce the current electron and positron data with a primary injection spectrum compatible with shock acceleration theory. Our model allows us to reproduce the AMS-02 PF accounting for an extra-component located in the Galactic arms with a high energy cutoff (10~TeV) and an harder spectrum. We also considered the possibility of different values of the cutoff, and we found that values down to 1~TeV yield a good fit the AMS-02 PF (see \cite{dragon3} for details). 

The physical interpretation of the extra component, according to the value of the cutoff, may be compatible either with an enhanced secondary production near the accelerator \cite{Blasi2009} or with a pulsar population located in the arms 
In the future we plan to exploit the anisotropic diffusion to produce even more realistic models of CR propagation. The role of anisotropic diffusion might be relevant in order to get a comprehensive model compatible both with CR electron and positron observations and with the observed $\gamma$-ray gradient (see \cite{gradient}). 
Also the diffusion in the local environment is expected to be highly anisotropic and our code is the ideal tool to study the impact of local structures on CR propagation.


\vspace{\baselineskip}

\end{document}